\documentclass[a4paper, 10pt, twocolumn]{article}

\usepackage[british]{babel}
\usepackage[utf8]{inputenc}
\usepackage[center]{titlesec}
\usepackage[margin = 1.6cm]{geometry}
\usepackage[labelsep = period, labelfont = bf, figurename = Figure, tablename = Table, font=small]{caption}

\usepackage{float}
\usepackage{times}
\usepackage{xcolor}
\usepackage{siunitx}
\usepackage{physics}
\usepackage{amsmath}
\usepackage{amssymb}
\usepackage{booktabs}
\usepackage{csquotes}
\usepackage{graphicx}
\usepackage{hyperref}
\usepackage{orcidlink}
\usepackage{tabularx}
\usepackage{makecell}
\usepackage{subcaption}
\usepackage{dblfloatfix}
\usepackage{indentfirst}
\usepackage{doi}
\usepackage{etoolbox}

\hypersetup{
    colorlinks = true,
    linkcolor = blue,
    urlcolor = blue,
    citecolor = blue,
    pdftitle = {Physics-Informed Neural Networks for Optimal Beam Shaping in Flat Optics}
}
\AtBeginEnvironment{thebibliography}{%
  \small
  \setlength{\itemsep}{0pt}
  \setlength{\parskip}{0pt}
  \setlength{\parsep}{0pt}
}

\titlespacing*{\section}{0pt}{5pt}{2pt}
\titlespacing*{\subsection}{0pt}{4pt}{1pt}

\begin{document}
    \makeatletter
    \twocolumn[
    \begin{@twocolumnfalse}
        \begin{center}
        {\LARGE \raggedright \textbf{Physics-Informed Neural Networks for Optimal Beam Shaping in Flat Optics}}\vspace*{4pt}
        
        {\noindent
        \rlap{\raisebox{-7pt}[0pt][0pt]{\textbf{Rafael de la Fuente Herrezuelo}}}%
        \phantom{\textbf{Rafael de la Fuente Herrezuelo}}
        }\vspace*{4pt}

        \phantom{July 2026}
        \end{center}

        \begin{abstract}
        We introduce a physics-informed neural network (PINN) approach for designing phase profiles in flat optics that reshape an incident beam into a prescribed target intensity distribution. The method solves Monge--Amp\`ere beam-shaping equations associated with energy-conserving ray mappings generated by a phase-only optical element. We treat both finite-distance and far-field targets using a generalized-Snell-law formulation. The learned phase profiles are validated by scalar diffraction simulations and compared with conventional phase-retrieval methods such as Gerchberg--Saxton. To our knowledge, this is the first time a PINN has been used for beam shaping problems in flat optics.
        \end{abstract}
        
        \noindent\rule{\textwidth}{.75pt}
    \end{@twocolumnfalse}
    ]
    \makeatother

\section{Introduction}

Beam shaping transforms an incident optical beam into a prescribed intensity distribution and is relevant for laser machining, illumination, holography, displays, and flat optics. In a phase-only implementation, an SLM or metasurface modifies the wavefront phase $\Phi(x,y)$ while leaving the input amplitude essentially unchanged, thereby redistributing the optical power after propagation.

Conventional phase-only beam shaping is often based on Fourier optics. The Gerchberg--Saxton (GS) algorithm alternates between the input and Fourier planes while replacing the amplitudes by the prescribed source and target amplitudes~\cite{Gerchberg1972}. GS is flexible for arbitrary far-field images, but the recovered phase is typically irregular and the reconstructed intensity may contain speckle-like fluctuations. A complementary route formulates beam shaping as an energy-conserving ray-mapping problem: generalized Snell law determines the ray direction from the local phase gradient, while conservation of optical power leads to a Monge--Amp\`ere equation for the phase~\cite{Fuente2025,NielsenThesis2024}.

Here we solve this Monge--Amp\`ere beam-shaping equation with a physics-informed neural network (PINN). This PINN approach is useful because the Monge--Amp\`ere equation is fully nonlinear and potentially degenerate, making standard finite-difference or finite-element solvers difficult to converge.  The network represents the phase potential for a fixed source--target pair and is trained by penalizing the PDE residual using automatic differentiation~\cite{Raissi2019}. This is not a data-driven surrogate: the training signal is the optical PDE itself. Related PINN ideas have been used in nano-optics and metamaterials~\cite{Chen2020}; here they are applied to flat-optics beam shaping. The approach generalizes recent cylindrically symmetric metasurface beam-shaping demonstrations~\cite{Nielsen2024,Poulsen2025} to non-cylindrically symmetric targets.

\section{Theoretical Framework}

Consider a phase-only flat optical element in the $(x,y)$ plane. The Monge--Amp\`ere equation is derived from a ray map in which the phase gradient determines the outgoing direction and a Jacobian enforces power conservation. The resulting phase is then tested independently with scalar wave propagation. We assume a thin element that changes the local phase while leaving the amplitude essentially unchanged immediately after the surface.

The incident beam propagates along $+z$, and just after the mask the scalar field is approximated by
\begin{equation}
    U_{\mathrm{out}}(x,y)
    \approx
    U_{\mathrm{in}}(x,y)\exp[i\Phi(x,y)] .
    \label{eq:phase_mask}
\end{equation}
The input irradiance is
\begin{equation}
    I(x,y)
    =
    \frac{1}{2}c\,n_{\mathrm{in}}\epsilon_0\,
    U_{\mathrm{in}}(x,y)^2 ,
    \label{eq:input_intensity}
\end{equation}
and the input and target intensities are normalized to equal total power, as required by a phase-only redistribution of optical energy. This also removes an arbitrary global intensity scale from the transport equations.

\begin{figure}[!t]
\centering
\includegraphics[width=\columnwidth]{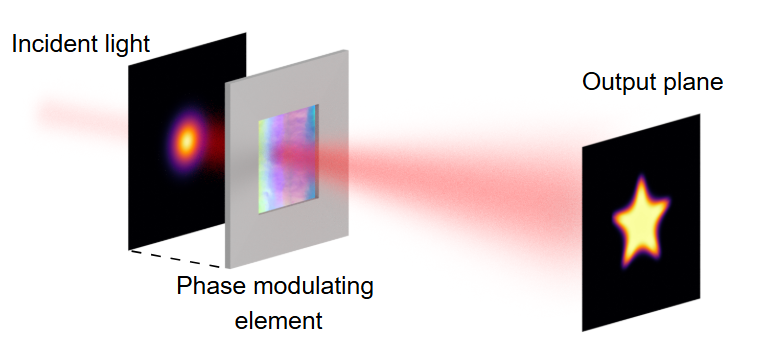}
\caption{Phase modulation for structured-beam generation. The incident field is phase-modulated by an SLM or metasurface and forms the target intensity after propagation.}
\label{fig:setup}
\end{figure}

\subsection{Generalized Snell law and ray mapping}
\label{subsec:snell}

The generalized Snell law~\cite{yu2011} relates the phase gradient to the tangential change of wave vector:
\begin{align}
n_{\mathrm{out}}\sin\theta_{\mathrm{out}}\cos\varphi_{\mathrm{out}}
  - n_{\mathrm{in}}\sin\theta_{\mathrm{in}}\cos\varphi_{\mathrm{in}}
  &=
  \frac{\lambda}{2\pi}\,\partial_x\Phi,
  \label{eq:snellx}\\
n_{\mathrm{out}}\sin\theta_{\mathrm{out}}\sin\varphi_{\mathrm{out}}
  - n_{\mathrm{in}}\sin\theta_{\mathrm{in}}\sin\varphi_{\mathrm{in}}
  &=
  \frac{\lambda}{2\pi}\,\partial_y\Phi .
  \label{eq:snelly}
\end{align}
For normal incidence, $\theta_{\mathrm{in}}=0$. Introducing
\begin{equation}
    \tilde{\Phi}=\frac{\Phi}{k},
    \qquad
    k=\frac{2\pi n_{\mathrm{out}}}{\lambda},
\end{equation}
the outgoing direction satisfies
\begin{equation}
    O_x=\tilde{\Phi}_x,
    \qquad
    O_y=\tilde{\Phi}_y,
    \qquad
    O_z=K,
\end{equation}
where
\begin{equation}
    K=
    \sqrt{1-\tilde{\Phi}_x^2-\tilde{\Phi}_y^2}.
    \label{eq:K}
\end{equation}
The potential $\tilde{\Phi}$ has units of length, so its transverse derivatives are dimensionless direction cosines. The condition $\tilde{\Phi}_x^2+\tilde{\Phi}_y^2\leq 1$ keeps $K$ real and the ray propagating. Retaining $K$ rather than setting it to unity preserves the angular dependence beyond a small-angle approximation. This distinction is relevant for strongly deflected rays, for which the longitudinal component $O_z=K$ changes appreciably across the aperture and directly modifies the finite-distance mapping.

A point $(x,y)$ is mapped to a finite-distance target plane at distance $z$ by
\begin{align}
    t_x
    &=
    x+z\,\frac{O_x}{O_z}
    =
    x+z\,\frac{\tilde{\Phi}_x}{K},
    \label{eq:mappingx}\\
    t_y
    &=
    y+z\,\frac{O_y}{O_z}
    =
    y+z\,\frac{\tilde{\Phi}_y}{K}.
    \label{eq:mappingy}
\end{align}
The phase gradient sets each ray destination, while second derivatives determine how neighbouring rays expand, compress, or shear. Consequently, local power conservation produces a nonlinear second-order PDE. A smooth solution also defines a continuous family of neighbouring rays, which is important because folds or caustics would make several source points reach the same target location and invalidate a single-valued transport map.

\subsection{Finite-distance Monge--Amp\`ere equation}

\begin{figure}[!t]
\centering
\includegraphics[width=\columnwidth]{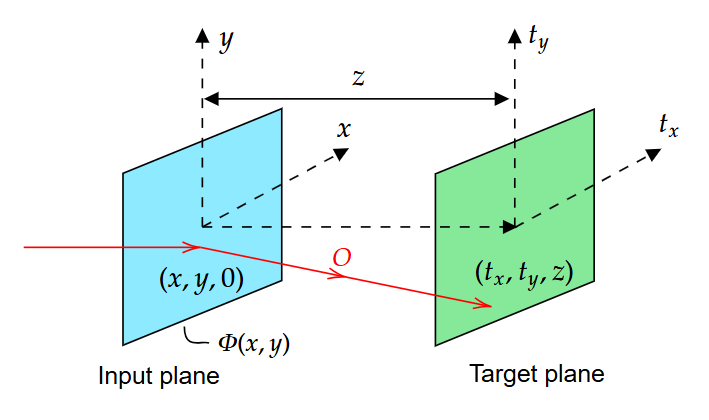}
\caption{Finite-distance geometry. The phase $\Phi(x,y)$ redirects rays from the input plane to target coordinates $(t_x,t_y)$.}
\label{fig:finite_distance}
\end{figure}

We enforce a ray map to deliver the correct power to every target region. Conservation of power between a corresponding source element $dx\,dy$ mapped to a target element $dt_x\,dt_y$  requires
\begin{equation}
    I(x,y)\,dx\,dy
    =
    E(t_x,t_y)\,dt_x\,dt_y .
    \label{eq:energybalance}
\end{equation}
Here $E(t_x,t_y)$ denotes the prescribed target intensity distribution and $I(x,y)$ the input irradiance. The Jacobian measures the local area change of the map, with expansion reducing irradiance and compression increasing it. Therefore,
\begin{equation}
    E(t_x,t_y)\,|J|=I(x,y),
    \label{eq:localbalance}
\end{equation}
where
\begin{equation}
    |J|
    =
    \left|
    \frac{\partial t_x}{\partial x}
    \frac{\partial t_y}{\partial y}
    -
    \frac{\partial t_x}{\partial y}
    \frac{\partial t_y}{\partial x}
    \right| .
    \label{eq:jacobian}
\end{equation}
Substitution of Eqs.~\eqref{eq:mappingx} and \eqref{eq:mappingy} gives
\begin{equation}
\begin{split}
\Big|
&A_1
\left(
\tilde{\Phi}_{xx}\tilde{\Phi}_{yy}
-\tilde{\Phi}_{xy}^{2}
\right)
+
A_2\,\tilde{\Phi}_{xx}
+
A_3\,\tilde{\Phi}_{yy}
\\
&+
A_4\,\tilde{\Phi}_{xy}
+
A_5
\Big|
=
K^4
\frac{I(x,y)}{E(t_x,t_y)} ,
\end{split}
\label{eq:MA_general}
\end{equation}
with
\begin{align}
    A_1 &= z^2, \\
    A_2 &= zK\left(1-\tilde{\Phi}_y^2\right), \\
    A_3 &= zK\left(1-\tilde{\Phi}_x^2\right), \\
    A_4 &= 2zK\,\tilde{\Phi}_x\tilde{\Phi}_y, \\
    A_5 &= K^4 .
\end{align}
The Hessian determinant in Eq.~\eqref{eq:MA_general} is the characteristic Monge--Amp\`ere nonlinearity, while the remaining terms arise from the finite-distance coordinates and the normalized slopes $\tilde{\Phi}_x/K$ and $\tilde{\Phi}_y/K$. Since $E$ is sampled at the phase-dependent coordinates $(t_x,t_y)$, the equation determines both the ray destinations and their local density. Admissible solutions keep the rays propagating and inside the target domain.

\begin{figure}[!t]
\centering
\includegraphics[width=\columnwidth]{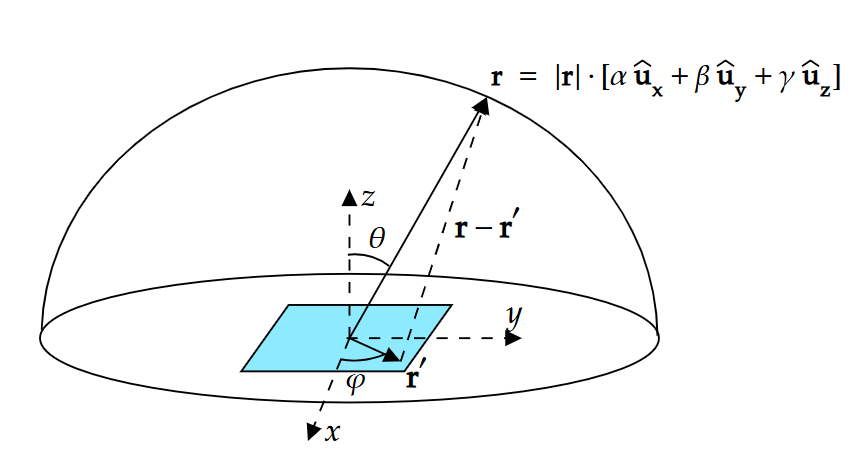}
\caption{Far-field geometry. The target is specified as an angular distribution $I_{e,\Omega\cos\theta}(\alpha,\beta)$.}
\label{fig:farfield}
\end{figure}

\begin{figure*}[!t]
\centering
\includegraphics[width=.9\linewidth]{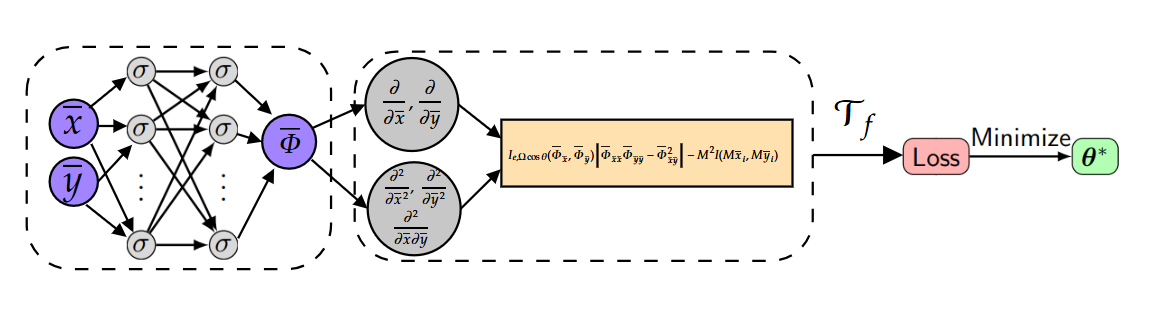}
\caption{PINN architecture. The network outputs $\bar{\Phi}$ and automatic differentiation supplies the derivatives entering the Monge--Amp\`ere residual.}
\label{fig:pinn_architecture}
\end{figure*}

\subsection{Far-field formulation}

For a finite target distance, each source point maps to a position in a plane. In the far field, the target is instead described by the radiant intensity per projected solid angle,
\begin{equation}
    \frac{\partial P_e}{\partial\Omega\cos\theta}
    = I_{e,\Omega\cos\theta},
    \label{eq:radiant_intensity}
\end{equation}
where $P_e$ is the emitted optical power. This distribution is parameterized by the direction cosines
\begin{equation}
\begin{split}
    \alpha &= \frac{x}{|\mathbf{r}|}
    =
    \sin\theta\cos\varphi,\\
    \beta &= \frac{y}{|\mathbf{r}|}
    =
    \sin\theta\sin\varphi,\\
    \gamma &= \frac{z}{|\mathbf{r}|}
    =
    \cos\theta
    =
    \sqrt{1-\alpha^2-\beta^2}.
\end{split}
\label{eq:dir_cosines}
\end{equation}
Because $d\alpha\,d\beta=\cos\theta\,d\Omega$, $I_{e,\Omega\cos\theta}$ can be represented directly in the $(\alpha,\beta)$ plane.

The mapping is simply
\begin{equation}
    \alpha=O_x=\tilde{\Phi}_x,
    \qquad
    \beta=O_y=\tilde{\Phi}_y .
    \label{eq:farfield_map}
\end{equation}
No propagation distance appears in Eq.~\eqref{eq:farfield_map}: the transverse ray components are already the target coordinates. Hence the far-field map is the gradient of $\tilde{\Phi}$ and its Jacobian is the Hessian determinant.

Conservation of optical power gives
\begin{equation}
\begin{split}
I(x,y)\,dx\,dy
&=
I_{e,\Omega\cos\theta}(\alpha,\beta)\,
d\alpha\,d\beta \\
&=
I_{e,\Omega\cos\theta}(\alpha,\beta)
\left|J(\mathbf{T}_{\alpha,\beta})\right|
dx\,dy .
\end{split}
\label{eq:farfield_balance_int}
\end{equation}
Since
\begin{equation}
    \left|J(\mathbf{T}_{\alpha,\beta})\right|
    =
    \left|
    \tilde{\Phi}_{xx}\tilde{\Phi}_{yy}
    -
    \tilde{\Phi}_{xy}^2
    \right|,
    \label{eq:farfield_jacobian}
\end{equation}
the far-field Monge--Amp\`ere equation is
\begin{equation}
\begin{split}
&I_{e,\Omega\cos\theta}
\left(
\tilde{\Phi}_x,\tilde{\Phi}_y
\right)
\left|
\tilde{\Phi}_{xx}\tilde{\Phi}_{yy}
-
\tilde{\Phi}_{xy}^2
\right|
\\
&\hspace{2.5cm}
=
I(x,y).
\end{split}
\label{eq:MA_farfield}
\end{equation}
Equation~\eqref{eq:MA_farfield} has the standard Monge--Amp\`ere structure of a gradient-generated transport map. The Hessian determinant controls the local area change, while the source and target densities determine the required expansion or compression. Compared with the finite-distance equation, the absence of the identity part of the map makes the far-field relation algebraically simpler: the target coordinate is the phase gradient itself.

The ray-based designs are validated independently with scalar diffraction simulations implemented using diffractsim~\cite{Diffractsim2022}.

\section{Physics-Informed Neural Network Framework}

\subsection{Normalized equations}

Direct optimization in physical units is poorly conditioned because coordinates, phase, and intensity can differ by several orders of magnitude. We therefore solve the PDE on a fixed dimensionless square with a phase scaling chosen to keep the ray map compact. The purpose of this normalization is not to alter the optics, but to present the optimizer with variables and derivatives of comparable magnitude.

For the finite-distance case, we use
\begin{equation}
    \bar{x}=\frac{x}{M},
    \qquad
    \bar{y}=\frac{y}{M},
    \qquad
    \bar{\Phi}=\frac{z}{kM^2}\Phi,
    \qquad
    s=\frac{M}{z}.
    \label{eq:scaling}
\end{equation}
Here $M$ is the aperture half-width and $s=M/z$. The factor $z/(kM^2)$ makes phase-gradient displacements naturally measured in units of $M$. Applying the chain rule gives
\begin{equation}
    \tilde{\Phi}_x=s\,\bar{\Phi}_{\bar{x}},
    \qquad
    \tilde{\Phi}_y=s\,\bar{\Phi}_{\bar{y}},
\end{equation}
and
\begin{equation}
    \bar{t}_x=
    \bar{x}
    +
    \frac{\bar{\Phi}_{\bar{x}}}{\bar{K}},
    \qquad
    \bar{t}_y=
    \bar{y}
    +
    \frac{\bar{\Phi}_{\bar{y}}}{\bar{K}},
    \label{eq:target_scaled}
\end{equation}
with
\begin{equation}
    \bar{K}
    =
    \sqrt{
    1-s^2
    \left(
    \bar{\Phi}_{\bar{x}}^2
    +
    \bar{\Phi}_{\bar{y}}^2
    \right)
    } .
    \label{eq:Kbar}
\end{equation}
The normalized finite-distance PDE reads
\begin{equation}
\begin{split}
\Big|
&\bar{A}_{1}
\left(
\bar{\Phi}_{\bar{x}\bar{x}}
\bar{\Phi}_{\bar{y}\bar{y}}
-
\bar{\Phi}_{\bar{x}\bar{y}}^{2}
\right)
+
\bar{A}_{2}\bar{\Phi}_{\bar{x}\bar{x}}
\\
&+
\bar{A}_{3}\bar{\Phi}_{\bar{y}\bar{y}}
+
\bar{A}_{4}\bar{\Phi}_{\bar{x}\bar{y}}
+
\bar{A}_{5}
\Big|
=
\bar{K}^{4}
\frac{
I(M\bar{x},M\bar{y})
}{
E(M\bar{t}_x,M\bar{t}_y)
}.
\end{split}
\label{eq:nearfield_bar}
\end{equation}
The coefficients are
\begin{align}
    \bar{A}_{1} &= 1,\\
    \bar{A}_{2} &=
    \bar{K}\left(1-s^2\bar{\Phi}_{\bar{y}}^2\right),\\
    \bar{A}_{3} &=
    \bar{K}\left(1-s^2\bar{\Phi}_{\bar{x}}^2\right),\\
    \bar{A}_{4} &=
    2s^2\bar{K}\bar{\Phi}_{\bar{x}}\bar{\Phi}_{\bar{y}},\\
    \bar{A}_{5} &= \bar{K}^{4}.
\end{align}
The transformation fixes the domain to $[-1,1]^2$ while retaining the nonlinear dependence on the gradient through $\bar{K}$. The target is still evaluated at the physical coordinates $(M\bar{t}_x,M\bar{t}_y)$, so normalization does not replace the prescribed physical irradiance profile. At each collocation point, the network and automatic differentiation provide the phase and its derivatives, after which the mapped target coordinate is updated and sampled. Because that coordinate changes during training, target evaluation is part of the differentiable computational graph rather than a fixed preprocessing step.

\begin{figure*}[!b]
\centering
\includegraphics[width=0.72\linewidth]{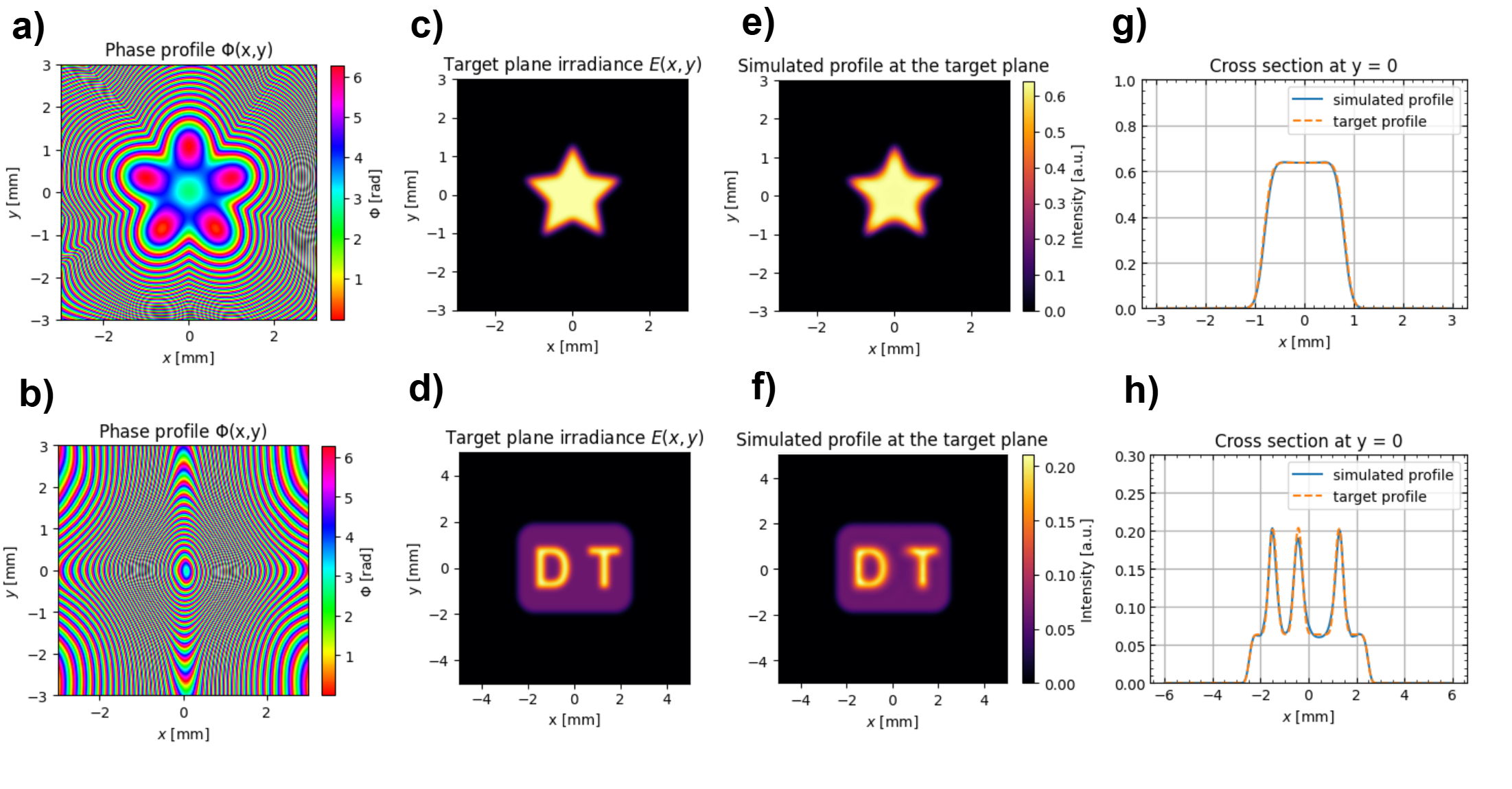}
\caption{PINN beam-shaping validation. The upper row shows a finite-distance Gaussian-to-star flat-top design. The lower row shows a far-field Gaussian-to-DT-logo design. The panels compare phase profiles, prescribed targets, propagated or far-field simulated irradiances, and central cross sections.}
\label{fig:finite_distance_results}
\end{figure*}

\begin{figure*}[!t]
\centering
\includegraphics[width=0.72\linewidth]{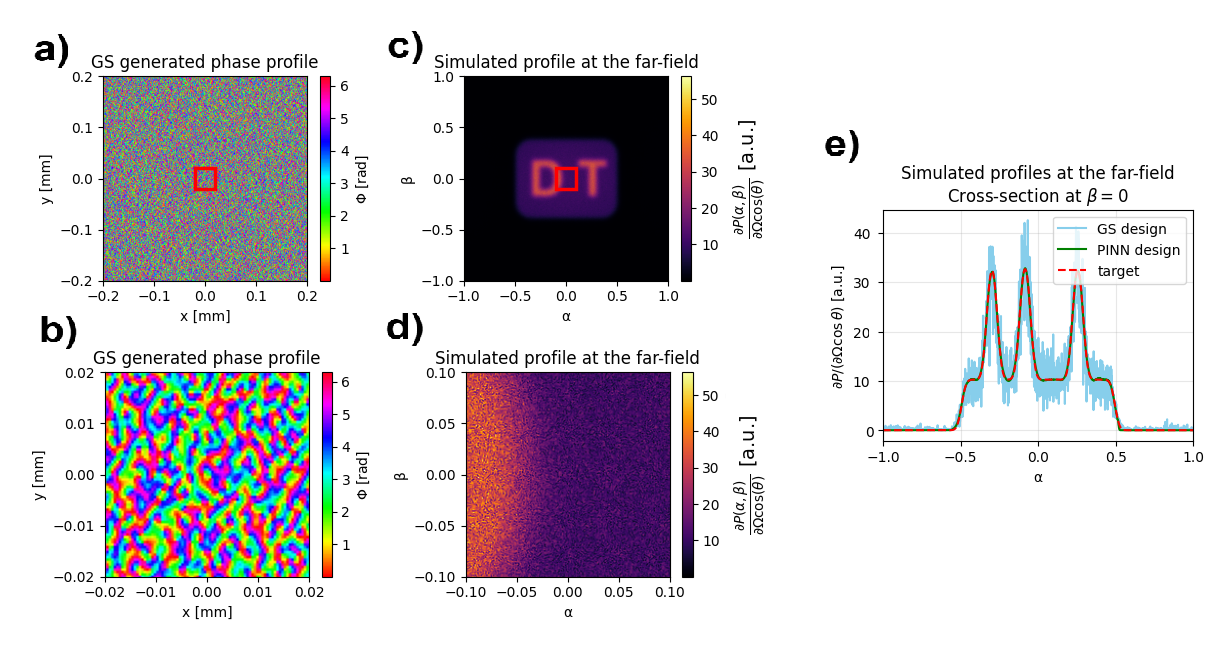}
\caption{Gerchberg--Saxton comparison for the DT far-field target. Panels (a,b) show the GS phase mask on the full aperture and in a central zoomed region (red box). Panels (c,d) show the corresponding far-field prediction on the full and zoomed angular domains. Panel (e) compares the $\beta=0$ cross sections of the GS prediction, the PINN prediction, and the target profile.}
\label{fig:gs_results_comparison}
\end{figure*}

For far-field targets, the scaling is instead
\begin{equation}
    \bar{x}=\frac{x}{M},
    \qquad
    \bar{y}=\frac{y}{M},
    \qquad
    \bar{\Phi}=\frac{\Phi}{kM}.
    \label{eq:scaling_far}
\end{equation}
Because the far-field target variables are directions rather than displacements, this scaling makes the direction cosines equal to derivatives with respect to the normalized aperture coordinates:
\begin{equation}
    (\alpha,\beta)
    =
    \left(
    \bar{\Phi}_{\bar{x}},
    \bar{\Phi}_{\bar{y}}
    \right),
\end{equation}
and the normalized far-field PDE is
\begin{equation}
\begin{split}
\left|
\bar{\Phi}_{\bar{x}\bar{x}}
\bar{\Phi}_{\bar{y}\bar{y}}
-
\bar{\Phi}_{\bar{x}\bar{y}}^{2}
\right|
=
M^2
\frac{
I(M\bar{x},M\bar{y})
}{
I_{e,\Omega\cos\theta}
\left(
\bar{\Phi}_{\bar{x}},
\bar{\Phi}_{\bar{y}}
\right)
}.
\end{split}
\label{eq:farfield_bar}
\end{equation}
The same network representation is used in both cases, but the residual retains the distinct physical maps: a finite-distance displacement or a far-field direction.

\subsection{PINN architecture}

For each source-target pair, the PINN represents one phase potential. It is a design-specific PDE solver, not a surrogate trained to predict phases for unseen targets. Training therefore replaces the unknown continuous phase function by a differentiable neural representation whose parameters are optimized for one prescribed transport problem.

The normalized phase is represented by a fully connected feedforward network,
\begin{equation}
    \bar{\Phi}(\bar{x},\bar{y})
    =
    \frac{1}{\sigma_{\Phi}}\,
    \mathcal{N}_{\theta}(\bar{x},\bar{y}),
    \label{eq:NNtophi}
\end{equation}
with two inputs, one scalar output, smooth $\tanh$ activations, and a linear output layer. Smooth activations are required because the residual contains second derivatives. We use four hidden layers with 150 neurons each.

The PyTorch implementation uses automatic differentiation for the first and second derivatives entering the Hessian. This avoids finite-difference stencils and keeps all derivative terms consistent with the same network representation. The resulting mapped coordinates depend on the trainable phase, so the target intensity is sampled with differentiable bilinear resampling, as in Spatial Transformer Networks~\cite{Jaderberg2015}. Gradients can therefore propagate through both the optical mapping and the interpolation of the target distribution.

The factor $\sigma_{\Phi}$ controls the typical phase-gradient scale. Values that are too small limit the reachable target displacement, while values that are too large can violate the propagating-ray condition. We select it by Bayesian optimization over validation loss and map admissibility.

The trainable parameters $\theta$ minimize the pointwise violation of power conservation. We multiply the determinant term by the target intensity to avoid division by small target values. For each collocation point $(\bar{x}_i,\bar{y}_i)$, we use
\begin{equation}
F_{\mathrm{near}}
=
E(M\bar{t}_x,M\bar{t}_y)
\left|\mathcal{D}_{\mathrm{near}}\right|
-
\bar{K}^{4}
I(M\bar{x}_i,M\bar{y}_i),
\end{equation}
where $\mathcal{D}_{\mathrm{near}}$ is the left-hand side of Eq.~\eqref{eq:nearfield_bar}. For the far-field case,
\begin{equation}
\begin{split}
F_{\mathrm{far}}
&=
I_{e,\Omega\cos\theta}
\left(
\bar{\Phi}_{\bar{x}},
\bar{\Phi}_{\bar{y}}
\right)
\left|
\bar{\Phi}_{\bar{x}\bar{x}}
\bar{\Phi}_{\bar{y}\bar{y}}
-
\bar{\Phi}_{\bar{x}\bar{y}}^{2}
\right|
\\
&\quad
-
M^{2}I(M\bar{x}_i,M\bar{y}_i).
\end{split}
\end{equation}
The loss is
\begin{equation}
    \mathcal{L}(\theta)
    =
    \frac{1}{N_c}
    \sum_{i=1}^{N_c}
    F(\bar{x}_i,\bar{y}_i;\theta)^2 .
    \label{eq:loss}
\end{equation}
All derivatives come from the same scalar network output. No phase labels or precomputed ray correspondences are used, so the training signal is entirely physics based. The collocation points only specify where the conservation law is tested; they do not provide examples of the desired phase.

Optimization uses Adam with stochastic mini-batches, followed by full-batch L-BFGS once the loss stabilizes. The second stage refines the coupled first- and second-derivative residual near sharp target features.

No Dirichlet condition is imposed on the phase value because an additive constant does not change the ray map. Physical admissibility is checked by requiring mapped points to remain in the target domain and $\bar{K}$ to remain real.

After convergence, the network is sampled on a regular grid and the resulting phase is evaluated with independent diffraction simulations. This separation between training and validation is important: the PINN is optimized with the ray-based conservation equation, whereas the reported beam-shaping performance is measured with wave propagation.

\section{Numerical Experiments}

\subsection{Gaussian to top-hat at a finite distance}

The finite-distance test reshapes a Gaussian beam into a smoothed star-shaped flat-top pattern. The source and target windows have side \SI{6}{mm}; the Gaussian waist is $w_I=\SI{1}{mm}$, the wavelength is $\lambda=\SI{1}{\micro\metre}$, and the target plane is placed at $z=\SI{10}{cm}$. The phase is represented by a four-hidden-layer $\tanh$ PINN with 150 neurons per layer, trained with the normalized near-field Monge--Amp\`ere residual and validated independently with scalar angular-spectrum propagation on a $2048\times2048$ grid.

The upper row of Fig.~\ref{fig:finite_distance_results} shows that the learned phase produces the desired star-like support after propagation. The target and simulated central cross sections nearly overlap, including the flat region and steep edges. Residual discrepancies are concentrated near the boundary, where finite-aperture diffraction and target smoothing limit the achievable sharpness.

\subsection{Gaussian to logo in the far field}

The far-field case uses the same fully connected PINN architecture with a Gaussian source over a \SI{1}{mm} aperture and a DT-letter target specified in direction-cosine coordinates $(\alpha,\beta)$. The target radiant intensity $I_{e,\Omega\cos\theta}(\alpha,\beta)$ is obtained from a smoothed image over $[-1,1]\times[-1,1]$. The lower row of Fig.~\ref{fig:finite_distance_results} shows that the simulated far-field intensity reproduces the rectangular DT support and the three main peaks in the $\beta=0$ cross section. This target is more demanding than the star because it combines a flat-top background with thin internal strokes, yet the PINN still yields a smooth phase potential and the correct global profile.

\subsection{Comparison with existing methods}

The DT far-field design was also computed with GS using the same input and target amplitudes. We used 60 iterations of the standard projection strategy~\cite{Gerchberg1972,Fuente2025}.

As shown in Fig.~\ref{fig:gs_results_comparison}, GS recovers the DT logo, but its irregular phase produces speckle and stronger cross-section oscillations. The PINN curve follows the target peaks more cleanly and suppresses much of this high-frequency structure.

For a quantitative comparison, both predictions were evaluated on $(\alpha,\beta)\in[-1,1]^2$ and rescaled by the least-squares scalar that best matches the target. The mean-squared error is approximately $3.59\times10^{6}$ for the PINN and $2.08\times10^{8}$ for GS; the relative root-mean-square errors are about $2.97\%$ and $22.6\%$, respectively. The energy efficiency,
\begin{equation}
    \eta =
    \frac{\text{energy flux inside the target support}}
         {\text{incident energy flux}},
\end{equation}
is also higher for the PINN: about $99.98\%$, compared with $95.90\%$ for GS. The PINN therefore provides greater accuracy and more target-directed power in this example.

\section{Conclusion}

We presented a PINN method for phase-only beam shaping in flat optics. The network represents the phase profile and is trained by minimizing the Monge--Amp\`ere residual associated with generalized-Snell-law ray mappings. The method applies to finite-distance targets and far-field angular targets, and extends previous cylindrically symmetric metasurface beam-shaping approaches to non-cylindrically symmetric patterns.

The finite-distance star example remains accurate under scalar angular-spectrum propagation, while the far-field DT-logo example demonstrates performance on a structured two-dimensional angular target. Compared with GS on the same DT target, the PINN produces less speckle, lower mean-squared error, lower relative RMS error, and higher power inside the target support. The agreement with independent diffraction simulations also confirms that the learned ray maps remain effective when wave effects are included. Therefore, the PINN provides superior accuracy and fidelity compared with GS for the tested flat-optics beam-shaping task.

\section*{Data availability}

All code and data is available at \url{https://github.com/rafael-fuente/pinn-shaper}.

\end{document}